\documentclass[epj]{svjour}
\usepackage{graphicx}
\begin{document}
\title{Nuclear collective dynamics within Vlasov approach}
\author{V. Baran \inst{1}, M. Colonna \inst{2}, M. Di Toro \inst{3}, B. Frecus \inst{1}, A. Croitoru \inst{1}, D. Dumitru \inst{1}  
}                     

\institute{Faculty of Physics, University of Bucharest, Romania \and Laboratori Nazionali del Sud, INFN, I-95123 Catania, Italy
\and Physics and Astronomy Department, University of Catania, Italy}
\date{Received: date / Revised version: date}
%
\abstract{
We discuss, in an investigation based on Vlasov equation, the properties of the
isovector modes in nuclear matter and atomic nuclei in relation with the symmetry energy. 
We obtain numerically the dipole response and determine the strength function for various systems, including
a chain of Sn isotopes. We consider for the symmetry energy three parametrizations with density 
providing similar values at saturation but which manifest very different slopes around this point.
In this way we can explore how the slope affects the collective response of finite nuclear systems.
We focus first on the dipole polarizability and show that while the  model is able to describe the expected mass dependence, $A^{5/3}$,
it also demonstrates that this quantity is sensitive to the slope
parameter of the symmetry energy. Then, by considering the Sn isotopic chain, we investigate the emergence of a collective mode,
the Pygmy Dipole Resonance (PDR), when the number of neutrons in excess increases. We show that the total energy-weighted sum rule exhausted by this mode has
a linear dependence with the square of isospin  $I=(N-Z)/A$, again sensitive to the slope of the symmetry energy with density. Therefore the
polarization effects in the isovector density have to play an important role in the dynamics of PDR.
These results provide additional hints in the investigations aiming to extract the properties of symmetry energy below saturation.
\PACS{
      {PACS-key}{21.65.Ef, 24.10.Cn, 24.30.Cz, 25.20.Dc }   
     } 
} 
\maketitle
\section{Introduction}
\label{intro}
After the early applications  
to the study of collisionless stellar dynamics \cite{jeans1915} and
in the context of plasma physics \cite{vlasov1938,bir1991}, 
the Vlasov equation has also proved to be very useful in the investigation
of the fermionic systems too, as are the atomic nuclei \cite{briNPA1981,berPR1988,bonPR1994,barPR2005}, the electrons
in atomic clusters \cite{calAP1997,fenRMP2010} or ultrafast electrons in thin metal films \cite{manPRB2004,croPRB2008}.
For these quantum systems the Vlasov equation appears as a semiclassical limit of the quantum dynamics.
In the field of nuclear physics, in many situations, the full quantum treatment is very involved and sometimes
a clear physical picture can be difficult to extract. The passage to a semi-classical
approach leads to simplifications and can provide a more transparent interpretation
of the studied processes. Of course, the price to pay for this transition is the
missing of important quantum correlations. However, as a reverse of the medal,
such an approach may help to distinguish the role of various quantum effects.
In the context of nuclear physics the quantum dynamics
at the level of mean-field approximation is described by the time-dependent
Hartree-Fock (TDHF) equation for the one-body density matrix $\rho$:
\begin{equation}
i\hbar \frac{d \rho}{dt}=[h_{MF}(\rho),\rho] \; ,
\label{tdhf}
\end{equation}
where $h_{MF}(\rho)$ is the single-particle Hartree-Fock Hamiltonian.
Working with the density matrix in the position representation, 
$ \rho_{\bf r'r''} $, the Wigner transform defined as a kind of Fourier transform, is introduced:
\begin{eqnarray}                                            
   \label{wigner_r}
 f({\bf r},{\bf p}) 
  \equiv \frac{1}{(2 \pi \hbar)^3} \, \rho_{\mathrm{Wigner}}({\bf r},{\bf p}) \nonumber \\
  = \frac{1}{(2 \pi \hbar)^3} 
     \int \! \mathrm{d}^3 s \; 
      \mathrm{e}^{- \frac{i}{\hbar} {\bf p}\cdot {\bf s}} \; 
      \rho_{{\bf r} + {\bf s}/2, {\bf r} - {\bf s}/2} \; .
\end{eqnarray}  
The exact equation satisfied by $f({\bf r},{\bf p},t)$, equivalent with TDHF, is:
\begin{eqnarray}                                            
   \label{moyal}
 \frac{\partial f({\bf r}, {\bf p}, t)}{\partial t} =
\frac{2}{\hbar}  
 sin \frac{\hbar}{2} (\nabla_{\bf r_1}\nabla_{\bf p_2}-\nabla_{\bf r_2}\nabla_{\bf p_1}) \nonumber \\
 h({\bf r_1},{\bf p_1})f({\bf r_2}, {\bf p_2}, t) |^{{\bf r_1}={\bf r_2}={\bf r}}_{{\bf p_1}={\bf p_2}={\bf p}}
 \; .
\end{eqnarray}
where $ \displaystyle h({\bf r},{\bf p})$ is the Wigner transformation of the  $h_{MF}(\rho)$.
This corresponds to a formulation of quantum mechanics in phase-space \cite{moyPCPS1949}, where the dynamics
is described in terms of Moyal brackets instead of Poisson brackets, and represents an appropriate
starting point for modeling quantum plasmas \cite{manFIC2005}. 
In the limit  $\hbar \rightarrow 0$, the Moyal bracket reduces to Poisson bracket and 
for a self-consistent mean-field approximated by a local one, we obtain the Vlasov equation :
\begin{equation}                                            
 \label{vlasov}
 \frac{\partial f({\bf r}, {\bf p}, t)}{\partial t} 
  + \frac{\bf p}{m} \cdot \nabla_{\bf r} f({\bf r},{\bf p},t) 
  - \nabla_{\bf r} U({\bf r}) \cdot \nabla_{\bf p} f({\bf r},{\bf p},t) 
  = 0  \; .
\end{equation}
In  the semi-classical limit the quantum effects
manifest through the Pauli correlations contained in the distribution function.
These are included already from the initial conditions, since the
Fermi-Dirac statistics should characterize the
initial distribution of the fermions. 

The purpose of this work is to describe recent
applications of this equation to the study of collective features in 
two-components systems by focusing on some new properties
of the Pygmy Dipole Resonance, a collective motion evidenced experimentally in nuclei with neutrons in excess.
For a minimal self-consistency of the presentation, we shall start with 
a description of some known but fundamental results concerning the Vlasov dynamics in nuclear systems.
This allows us to introduce the basic
definitions and concepts required to characterize the nuclear systems and also will provide a reference for
the results obtained later numerically. Then we describe the numerical implementation of the Vlasov equation to the
nuclear dynamics and study in detail the dipole response by employing three different parametrization with density of
the symmetry energy. We determine the strength function and investigate the mass and isospin dependence of
the polarizability and of the sum-rule exhausted by PDR. 

\section{Isovector modes in nuclear matter}
\label{sec:1}
In the present section we describe the Vlasov approach to the
collective dynamics of nuclear matter. This is a two component system 
and therefore the semi-classical dynamics is determined by two coupled Vlasov equations:
\begin{eqnarray}                           
 \frac{\partial f_p}{\partial t} + \frac{\bf p}{m} \cdot \nabla_{\bf r} f_p 
  - \nabla_{\bf r} U_p \cdot \nabla_{\bf p} f_p 
  &= 0  \; , \label{vlaprot} \\
 \frac{\partial f_n}{\partial t} + \frac{\bf p}{m} \cdot \nabla_{\bf r} f_n 
  - \nabla_{\bf r} U_n \cdot \nabla_{\bf p} f_n 
  &= 0  \; \label{vlaneut}.
\end{eqnarray}
The nuclear mean-field contains an isoscalar as well as an isovector part
and here we adopt a Skyrme-like ($SKM^*$) parametrization with nucleons density $\rho=\rho_n+\rho_p$:
\begin{equation}
U_{q} = A\frac{\rho}{\rho_0}+B(\frac{\rho}{\rho_0})^{\alpha+1} + C(\rho)
\frac{\rho_n-\rho_p}{\rho_0}\tau_q
+\frac{1}{2} \frac{\partial C}{\partial \rho} \frac{(\rho_n-\rho_p)^2}{\rho_0}~~~.
\label{meanfield}
\end{equation}
Here $\rho_n(\bf{r})$ ($\rho_p(\bf{r})$) is the neutron (proton) local  density and $\tau_n (\tau_p)=+1 (-1)$.
The saturation properties of the symmetric nuclear matter, the density $\rho_0=0.16 fm^{-3}$, the binding energy
$E/A=-16 MeV$ and a compressibility modulus $K=200 MeV$, are reproduced with the values for the coefficients
$A=-356 MeV$, $B=303 MeV$, $\alpha=1/6$. 
In the expression of total energy per particle, 
\begin{equation}
 \frac{E}{A}(\rho, I) =\frac{E}{A}(\rho) +\frac{E_{sym}}{A}(\rho) I^2
\end{equation}
the quantity depending on the isospin parameter $ \displaystyle I=\frac{N-Z}{A}$, 
define the symmetry energy of the system.
\begin{figure}
\begin{center}
\includegraphics*[scale=0.35]{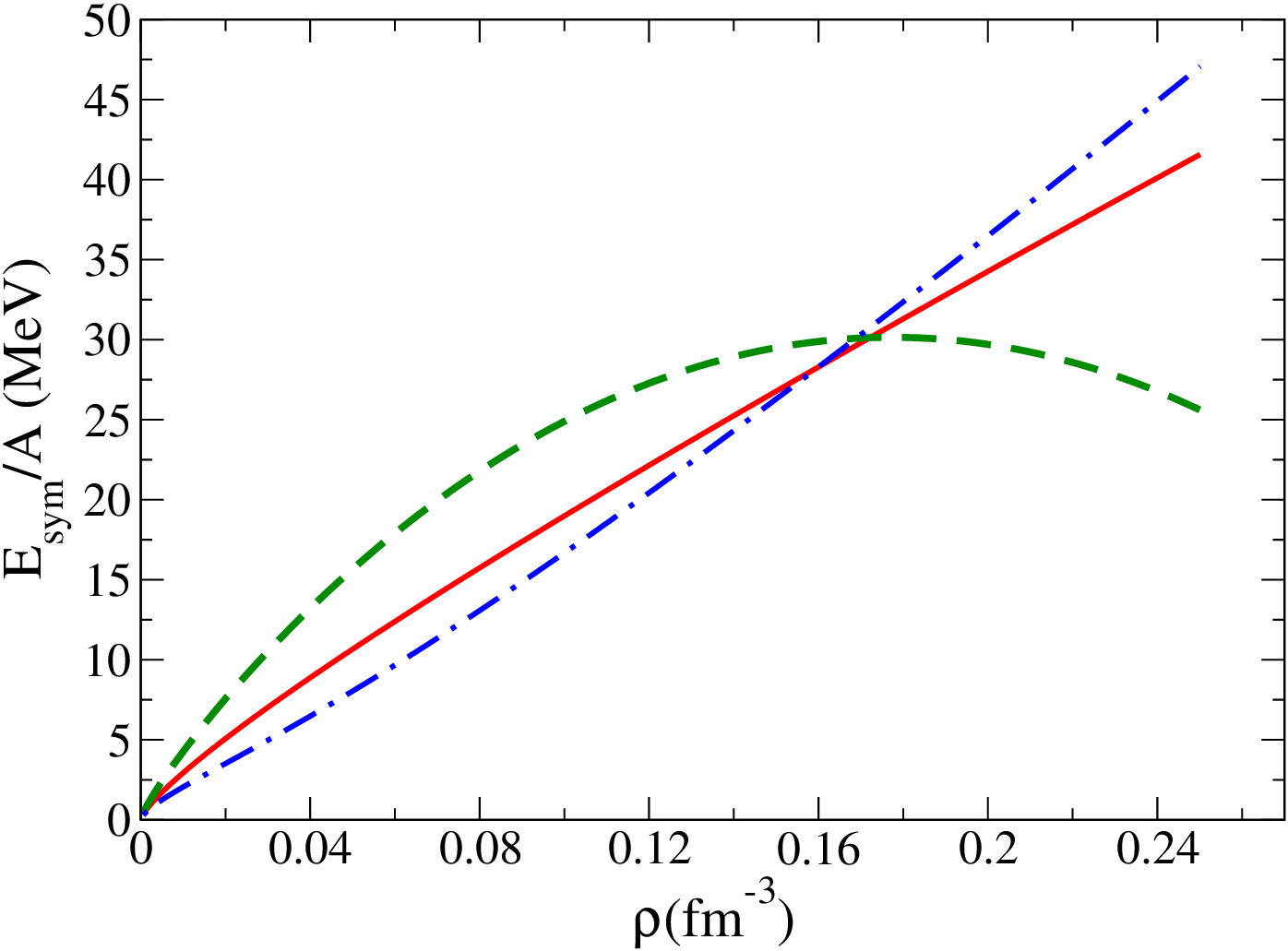}
\end{center}
\caption{(Color online) The density dependence of $\displaystyle \frac{E_{sym}}{A}$ for asysoft (dashed green line), asystiff (solid red line)
and asysuperstiff (dot-dashed blu line).}
\label{esymet}       
\end{figure}
Since the dipole response is determined by the symmetry energy term, containing both a kinetic and a potential contribution: 
\begin{equation}
 E_{sym}/A \equiv \epsilon_{sym}={E_F(\rho) \over 3}+{C(\rho) \over 2}{\rho \over \rho_0}
\end{equation}
with $E_F$ being the Fermi energy, we shall consider three different parametrizations with density of $C(\rho)$.
We selected those parametrizations which provide quite similar values of $E_{sym}/A$ at saturation but
manifest very different slopes around this point, see Fig. \ref{esymet}.
Specifically, for asystiff EOS, the coefficient $C(\rho)$ is constant,
 $ \displaystyle C(\rho) = 32MeV$. Then the symmetry energy 
 at
saturation is $E_{sym}/A = 28.3 MeV$ and the slope parameter 
$\displaystyle L = 3 \rho_0 \frac{d E_{sym}/A}{d \rho} |_{\rho=\rho_0}$
takes the value $L=72 MeV$. 
For the asysoft case we have: 
\begin{equation}
\frac{C(\rho)}{\rho_0} = (482-1638 \rho) MeV fm^3
\end{equation}
which leads to a quite small value 
of the slope parameter $L= 14.4 MeV$.
For the asysuperstiff EOS, 
\begin{equation}
\frac{C(\rho)}{\rho_0}=\frac{32}{\rho_0} \frac{2 \rho}{(\rho + \rho_0)}
\end{equation}
the symmetry term has a faster variation around saturation density with a slope parameter $L=96.6 MeV$.

To study the collective modes of nuclear matter we consider small deviations of the distribution functions from
equilibrium: 
\begin{eqnarray}                                        
 f_p({\bf r},{\bf p}, t) 
  = f^0_p(E) + \delta f_p({\bf r},{\bf p}, t)  \; ,  \\
 f_n({\bf r},{\bf p}, t) 
  = f^0_n(E) + \delta f_n({\bf r},{\bf p}, t)  \; ,
\end{eqnarray}
with $ \displaystyle f^0_p(E) = f^0_n(E) = \frac{\gamma}{(2 \pi \hbar)^3} \; \Theta(E - E_F) $
representing the equilibrium Fermi-Dirac distribution functions
for protons and neutrons and $\gamma=2$. Within the linear approximation, the
Eqs. (\ref{vlaprot}),(\ref{vlaneut}) become:
\begin{eqnarray}                                        
 \frac{\partial \, \delta f_p}{\partial t} 
  + \frac{\bf p}{m} \cdot \nabla_{\bf r} \delta f_p 
  - \nabla_{\bf r}\delta U_p \cdot \nabla_{\bf p}  f_p^0 
 = 0  \; , \\
 \frac{\partial \, \delta f_n}{\partial t} 
  + \frac{\bf p}{m} \cdot \nabla_{\bf r} \delta f_n 
  - \nabla_{\bf r} \delta U_n \cdot \nabla_{\bf p}  f_n^0 
 = 0  \; .
\end{eqnarray}
For the Skyrme-like parametrization of the mean-field potentials for
protons and neutrons (see  Eqs. (\ref{meanfield})) one has: 
\begin{eqnarray}                                        
 \delta U_p 
  = \frac{A}{\rho_0} (\delta \rho_n + \delta \rho_p ) 
     + \frac{ B}{\rho_0} (\delta \rho_n + \delta\rho_p)^\sigma 
     + \frac{C}{\rho_0} (\delta\rho_p - \delta \rho_n) \; , \nonumber \\
 \delta U_n 
  = \frac{A}{\rho_0} (\delta \rho_n + \delta \rho_p ) 
     + \frac{ B}{\rho_0} (\delta \rho_n + \delta\rho_p)^\sigma 
     + \frac{C}{\rho_0} (\delta\rho_n - \delta \rho_p) \; \nonumber .
\end{eqnarray} 
Then the isovector variation $ \delta f = \delta f_p - \delta f_n $, satisfy the equation:
\begin{equation}                                            
 \frac{\partial \, \delta f}{\partial t} 
  + \frac{\bf p}{m} \cdot \nabla_{\bf r} \delta f 
  - \nabla_{\bf r}(\delta U_p - \delta U_n) \cdot 
     \Big( - \frac {\bf p}{m} \; \delta (E-E_F) \Big) 
  = 0  \; ,
\label{isovec}
\end{equation}
where
\begin{equation}                                            
 (\delta U_p - \delta U_n) 
  = 2 \; \frac{C}{\rho_0} \; (\delta\rho_p - \delta \rho_n)
  = 2 \; \frac{C}{\rho_0} \int \! \mathrm{d}^3p \; \gamma \; \delta f \; .
\end{equation}
Searching for a plane-wave solution:
\begin{equation}                                            
 \delta f({\bf r},{\bf p}, t) 
  = \frac{\gamma}{(2 \pi \hbar)^3} 
     \sum_{\bf k} \, A_{\bf k}({\bf p}) \; \mathrm{e}^{\, i ({\bf k r} - \omega t)} \; ,
\end{equation}
one can arrive, with $ \displaystyle s = \frac{\omega}{k \, v_F}$, at the dispersion relation:
\begin{equation}
 \frac{s}{2} \, \ln \frac {s+1}{s-1} - 1 = \frac{2 \, E_F}{3 \, C}  \; .
 \label{dispers}
\end{equation}
The isovector collective mode velocity $\displaystyle \frac{\omega}{k}$ will depend now on the symmetry energy
through the value of the parameter $ \displaystyle C $. For $ \displaystyle C = 32 $ MeV the
solution is $ \displaystyle s = \frac{\omega}{k \, v_F} \approx 1.08 $. 
The isovector zero sound mode in nuclear matter is a quantum collective motion 
different from the hydrodynamical (or first sound) mode. While the first sound 
is driven by the pressure gradient, and so, requires the local thermodynamical
equilibrium, the zero sound is determined by the mean-field and therefore 
propagates even at $T=0 MeV$.  
In finite nuclei, this isovector zero sound mode will correspond to the Giant Dipole Resonance.
Considering a wave-number $ \displaystyle k $ determined by the size of the Tin isotope
$^{132}$Sn, the energy of the GDR phonon would be expected around
$\hbar \, \omega \approx 15 $ MeV, quite close to the experimental results. 
In the next section we investigate numerically the dynamics of this mode in finite
systems and study its evolution with the number of excess neutrons $\displaystyle N-Z$ by considering 
a chain of Sn isotopes, $^{108,116,124,132,140,148}Sn$. 

We would like complete this general presentation mentioning that
similar  investigations can be extended to the study of instabilities in nuclear
matter \cite{barPRL2001}. Moreover, if the present treatment is completed by adding  
a collision term in the relaxation time approximation, the nature of the zero to first sound
transition in binary Fermi liquids can be explored \cite{larNPA1999,barNPA1999}.

\section{Giant and Pygmy Dipole Resonances in neutron rich nuclei}
\label{sec:2}
\subsection{Numerical implementation and static properties}
The essential task of the transport approach based on Vlasov equation
is to provide the value of the one-body distribution function at any time. 
Once this quantity is known, the
expectation values of any one-body observable, $ \displaystyle A({\bf r},{\bf p}) $, can be
evaluated as an integral over phase-space. 
In particular, the total number of protons, $ Z $, is:
\begin{equation}                                            
Z = \int \mathrm{d}^3 {r} \; \mathrm{d}^3 {p} \; 
  f_p({\bf r}, {\bf p}, t) \; .
\end{equation}
The numerical approach employed in our work is based on test particle method. 
This method starts from the observation that the Gaussian functions
generate, as coherent states, a super-complete basis.  
Then, the following expansion for the distribution function is valid:
\begin{eqnarray}                                            
 f({\bf r},{\bf p},t) 
  = \frac{1}{(2\pi\hbar)^3} 
     \int \! \mathrm{d}^3 {r}_0 \; \mathrm{d}^3 {p}_0 \;  \nonumber \\
      \omega({\bf r}_0, {\bf p}_0, 0) \; 
      g({\bf r} - {\bf r}_0,{\bf p} - {\bf p}_0, t)  \; ,
\end{eqnarray}
where $ g({\bf r} - {\bf r}_0, {\bf p} - {\bf p}_0, t) $ corresponds to a 
product of Gaussian functions in coordinate and momentum space, centered in 
$ \displaystyle {\bf r}_0 $ and $ {\bf p}_0 $ respectively, while 
$ \displaystyle \omega({\bf r}_0, {\bf p}_0, 0) $ represents the corresponding weight in the
expansion of $ g $.  
In our numerical implementation, this expansion is discretized into a sum over a
sufficiently large number of Gaussian functions \cite{greNPA1987,schPPNP1989}: 
\begin{eqnarray}                                            
 f({\bf r},{\bf p},t)
  = \frac{1}{\mathcal{N}} \, \frac{1}{(2\pi\hbar)^3} \, 
     \frac{1}{(4 \pi^2 \chi \phi)^{3/2}} \nonumber \\
    \sum_i^N \exp \bigg(-\frac{({\bf r} - {\bf r}_i(t))^2}{2 \, \chi} \bigg) 
            \exp \bigg(-\frac{({\bf p} - {\bf p}_i(t))^2}{2 \, \phi} \bigg) \; .
 \label{f_test_particle}
\end{eqnarray}
Here $ \displaystyle {\bf r}_i(t) $ and $ \displaystyle {\bf p}_i(t) $ define the centroid positions in
coordinates and momentum space of the $ i $-Gaussian function, which corresponds to an
individual test particle. $ \displaystyle \mathcal{N} $ indicates the number of test
particles per nucleon. The total number of test particles will be $ \displaystyle N_{tot} = A\cdot\mathcal{N} $ 
and its value is limited by the requirement to have a reasonable computational effort.
From a comparison with the particle-in-cell (PIC) simulation method employed in plasma dynamics \cite{bir1991}, we observe that while
in a present treatment several "test particles" are ascribed to a real particle, in order to obtain a very good spanning of phase-space and reduce
the numerical fluctuations, in the PIC approach can be considered "finite size superparticles" which will incorporate
several real particles, reducing so the numerical effort, 
in such a way that charge density, the mass density and the thermal energy density match the same
values as those of the real particles and the same long-range behavior as in real plasma is maintained \cite{birPF1970}.  
The Gaussian functions which appear in Eq. (\ref{f_test_particle}) are 
normalized as follows: 
\begin{eqnarray}                                        
 g_{\chi}({\bf r} - {\bf r}_i) 
  = \frac{1}{(2 \pi \chi )^{3/2}} \; 
      \exp \bigg(-\frac{({\bf r} - {\bf r}_i)^2}{2\chi} \bigg) \; , \\
 g_{\phi}({\bf p} - {\bf p}_i) 
  = \frac{1}{(2 \pi \phi)^{3/2}} \; 
      \exp \bigg(-\frac{({\bf p} - {\bf p}_i)^2}{2\phi} \bigg) \; ,
\end{eqnarray}
so that one has:
\begin{equation}                                        
\int \! \mathrm{d}^3 {r} \; g_{\chi}({\bf r} - {\bf r}_i) 
  = 1  \; ;
\int \! \mathrm{d}^3 {p} \; g_{\phi}({\bf p} - {\bf p}_i) 
  = 1 \; .
\end{equation}
Substituting these expressions into the Vlasov equation (Eq. (\ref{vlasov})), we
find that the centroids of each Gaussian, $ {\bf r}_i $ and $ {\bf p}_i $ must
satisfy Hamilton equations, that is: 
\begin{eqnarray}    \label{hamilton_eq}                 
 \frac{ \partial {\bf r}_i}{\partial t} = \frac{{\bf p}_i}{m} \; , \\
  \frac{ \partial {\bf p}_i}{\partial t} = - \, \nabla_{{\bf r}_i} U({\bf r}_i) \; .
\end{eqnarray}
More generally, the average value of an arbitrary physical quantity, 
$A({\bf r},{\bf p})$, can be expressed:
\begin{eqnarray}                                             
 & \langle A({\bf r},{\bf p})\rangle 
   = \int \! \mathrm{d}^3 {r} \; \mathrm{d}^3 {p}   \; 
      A({\bf r},{\bf p}) f({\bf r},{\bf p},t) \nonumber  \\
   &= \frac{1}{\mathcal{N}} \, \frac{1}{(2\pi\hbar)^3} 
      \sum_i^N \int \! \mathrm{d}^3 {r} \; \mathrm{d}^3 {p} \; 
       A({\bf r},{\bf p}) \; g_{\chi}({\bf r} - {\bf r}_i) \; 
       g_{\phi}({\bf p} - {\bf p}_i) \nonumber \\
   &= \frac{1}{\mathcal{N}} \, \frac{1}{(2\pi\hbar)^3} 
      \sum_i^N \, \langle A({\bf r}, {\bf p}) \rangle_i \; , 
\end{eqnarray}
where $ \langle A({\bf r},{\bf p})\rangle_i $ represents the contribution of an
individual Gaussian. 
This contribution is obtained by the convolution of $ A $ with the corresponding
Gaussians in coordinate and momentum space: 
\begin{equation}                                            
\langle A({\bf r}, {\bf p}) \rangle_i 
= \int \! \mathrm{d}^3 {r} \; \mathrm{d}^3 {p} \; 
A({\bf r}, {\bf p}) \; g_{\chi}({\bf r} - {\bf r}_i(t)) \; 
g_{\phi}({\bf p} - {\bf p}_i(t)) \; .
\end{equation}
The method described above can reproduce accurately the equation of state of the nuclear matter as well as
the properties of the nuclear surface \cite{idiNPA1993} and the ground state energy
for finite nuclei \cite{schPPNP1989}. Here we discuss the predictions concerning the neutron
skin of the Sn isotopic chain when a number of
$1300$ t.p. per nucleon is adopted. From the one-body distribution functions one obtains the local densities: 
$\displaystyle \rho_q(\vec{r},t)=\int 2 d^3 {\bf p} f_q(\vec{r},\vec{p},t)$
as well as the quadratic radii
$\displaystyle \langle r_q^2 \rangle = \frac{1}{N_q} \int r^2 \rho_q(\vec{r},t) d^3 {\bf r}$
and the width of the neutrons skin
$\displaystyle \Delta R_{np}= \sqrt{\langle r_n^2 \rangle}-\sqrt{\langle r_p^2 \rangle}=R_n-R_p$.
A possible method to obtain $\displaystyle R_n$ and $\displaystyle R_p$
is by observing their time evolution after a weak monopolar perturbation.
Both quantities perform small oscillations
around equilibrium values and we remark that the numerical simulations keep
a very good stability of the dynamics for at least $1800 fm/c$ \cite{barPRC2013}.
Using this procedure we obtain for the charge mean square radius of $^{208}Pb$ a value around 
$\displaystyle R_p=5.45 fm$, to be compared with
the experimental value $\displaystyle R_{p, exp}=5.50 fm$. 
Analogously for $^{124}Sn$ we obtain $\displaystyle R_p=4.59 fm$ while experimentally $\displaystyle R_{p, exp}=4.67 fm$.
\begin{figure}
\begin{center}
\includegraphics*[scale=0.42]{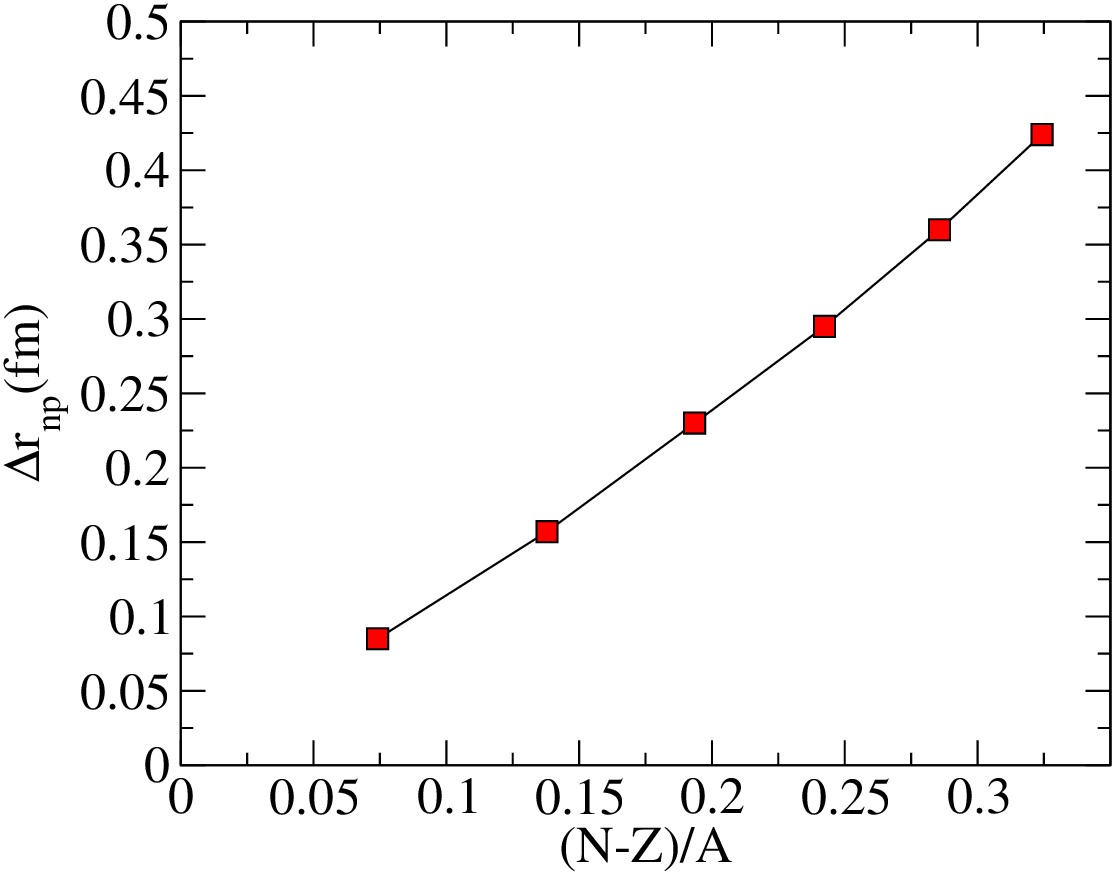}
\end{center}
\caption{(Color online) The neutron skin thickness as a function of $ \displaystyle I=\frac{N-Z}{A}$ for 
$\displaystyle ^{108,116,124,132,140,148}Sn$ isotopes. Asystiff EOS.}
\label{skin}       
\end{figure} 
For Sn isotopes and asystiff EOS we display the isospin parameter $\displaystyle I=\frac{N-Z}{A}$ dependence of 
$\displaystyle \Delta R_{np}$ respectively in Fig. \ref{skin}.

\subsection{Collective dipole response: the emergence of Pygmy Dipole Resonance}
We explore in this section, for various nuclear systems, some new the features of the E1 response in the Vlasov
approach. We consider a GDR-like initial
condition \cite{barPRC2012} which corresponds to a boost of all neutrons against all protons 
while keeping the Center of Mass (CM) at rest. This is described by the instantaneous
excitation $\displaystyle V_{ext} =\eta \delta(t-t_0) \hat{D}$ at $t=t_0$ \cite{calAP1997}. 
\begin{figure}
\begin{center}
\includegraphics*[scale=0.4]{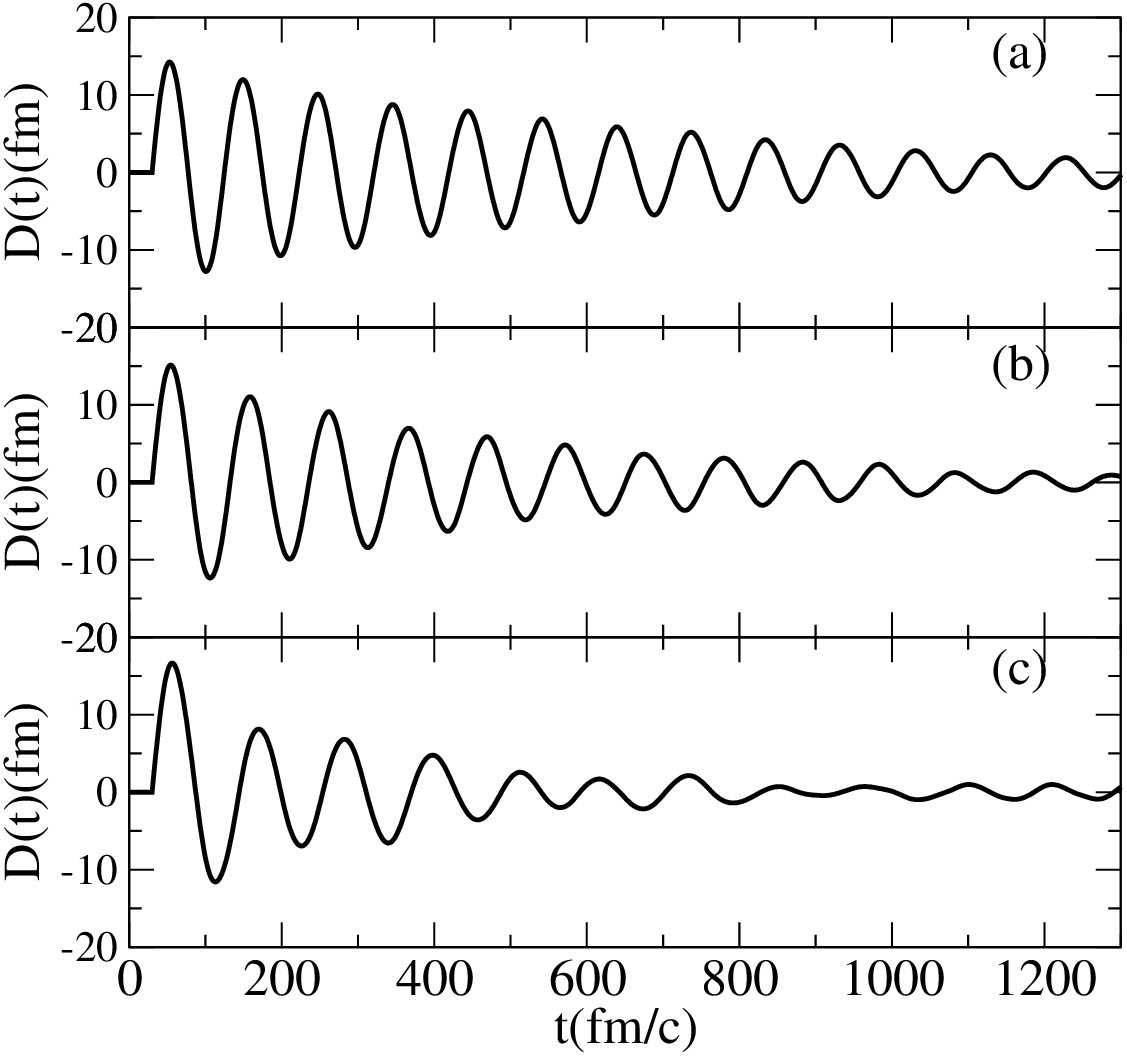}
\end{center}
\caption{The time evolution of the dipole moment following a GDR-like initial condition
for (a) $^{108}Sn$, (b) $^{124}Sn$ and (c) $^{148}Sn$. Asystiff EOS.}
\label{dip}       
\end{figure}
If $|\Phi_{0} \rangle$ is the state before perturbation then the excited state 
becomes $\displaystyle |\Phi (t_0)\rangle 
=e^{i \eta \hat{D}} |\Phi_{0} \rangle$, where $\hat{D}$ is the dipole operator.
The value of $\eta$ can be related to the initial 
expectation value of the collective dipole momentum $\hat{\Pi}$:
\begin{equation}
\langle \Phi (t_0)|\hat{\Pi}|\Phi (t_0) \rangle  = \hbar \eta \frac{N Z}{A}.
\label{eta}
\end{equation}
If the collective coordinate  which defines the 
distance between the CM of protons and the CM of neutrons is $\hat{X}$, 
then $\hat{\Pi}$ is canonically conjugated momentum, i.e. $[\hat{X},\hat{\Pi}]=i\hbar$ \cite{barRJP2012}.
We determine semi-classically the strength function:
\begin{equation}
S(E)=\sum_{n > 0}|\langle n|\hat{D}|0\rangle|^2\delta(E-(E_n-E_0)) ,
\end{equation} 
where $E_n$ are the excitation energies of the states $|n\rangle$ while
$E_0$ is the energy of the ground state $\displaystyle |0\rangle=|\Phi_{0} \rangle$. In our approach this is obtained  
from the imaginary part of the Fourier transform of the time-dependent expectation value of 
the dipole momentum $ \displaystyle D(t) = \frac{NZ}{A} X(t)= \langle \Phi (t) |\hat{D}| \Phi (t) \rangle $ extracted
from our simulations (see Fig. \ref{dip}). We have:
\begin{equation}
 S(E) =\frac{Im(D(\omega))}{\pi \eta \hbar}~~,
\label{stre}
\end{equation}
where $\displaystyle D(\omega) =\int_{t_0}^{t_{max}} D(t) e^{i\omega t} dt$.
We consider the initial perturbation along the z-axis and integrate numerically
the Vlasov equations (\ref{vlaprot}, \ref{vlaneut}) until $t_{max}=1830fm/c$.
$\eta$ was determined from the numerical value of the collective momentum at $t=t_0=30fm/c$. 
\begin{figure}
\begin{center}
\includegraphics*[scale=0.36]{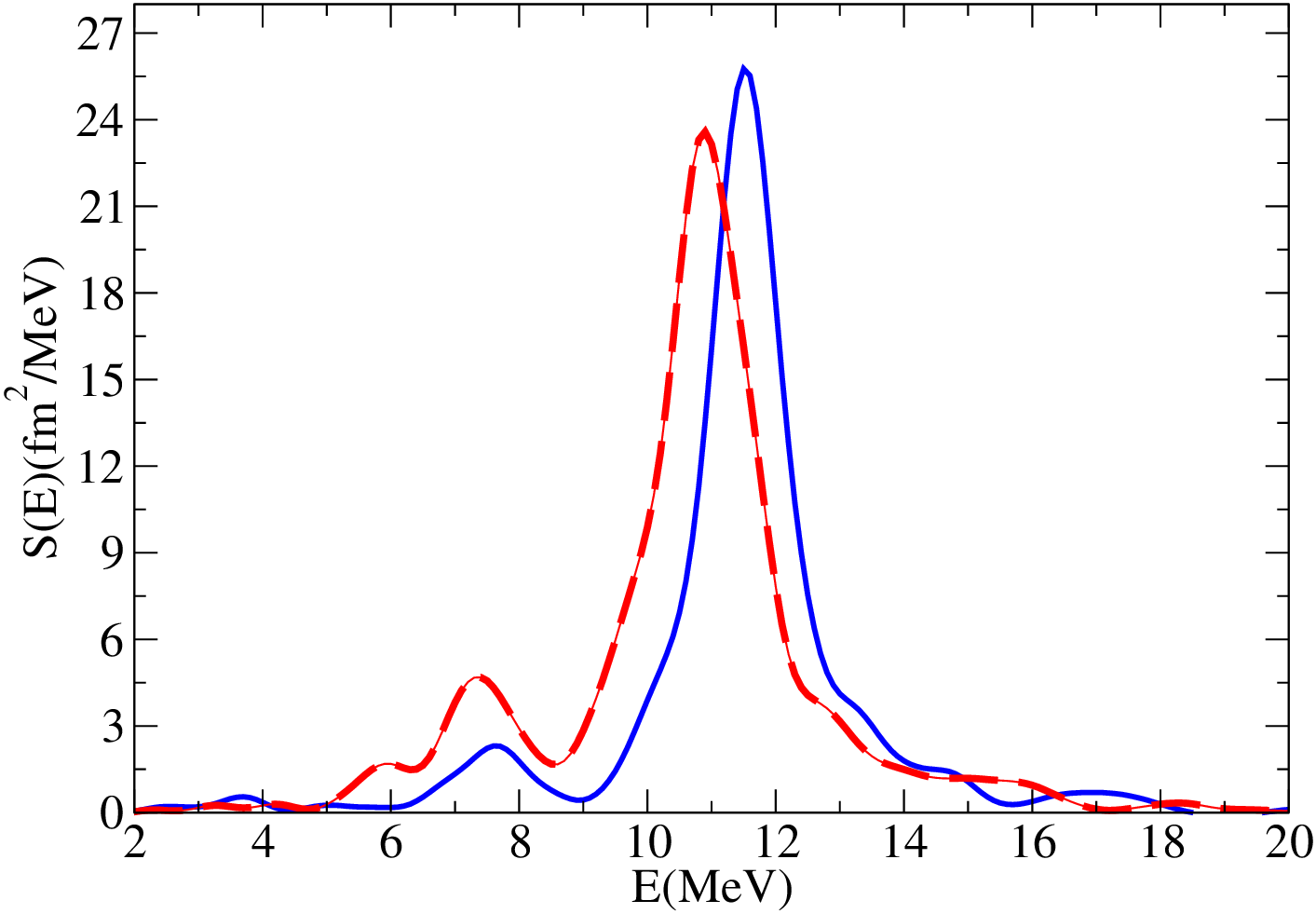}
\end{center}
\caption{(Color online) The strength function for $^{132}Sn$ [the blue (solid) lines] and $^{148}Sn$ [the red (dashed) lines].
Asystiff EOS.}
\label{stre}       
\end{figure} 
In order to eliminate the artifacts resulting from a finite time domain analysis of the signal a filtering procedure, as described
in \cite{reiPRE2006}, was considered. A smooth cut-off function was introduced such
that $D(t) \rightarrow D(t)cos^{2}(\frac{\pi t}{2 t_{max}}) $. 
The E1 strength functions of $^{132}Sn$ and $^{148}Sn$ 
are represented in Fig. \ref{stre}. A test of the quality of our method is the comparison of the  numerically estimated
value of the first moment $\displaystyle m_1=\int_0^\infty E S(E) dE$  with the value predicted by the
Thomas-Reiche-Kuhn (TRK) sum rule $\displaystyle m_1= \frac{\hbar^2}{2m} \frac{N Z}{A}$. In all cases 
the difference was below $5\%$. 

\begin{figure}
\begin{center}
\includegraphics*[scale=0.36]{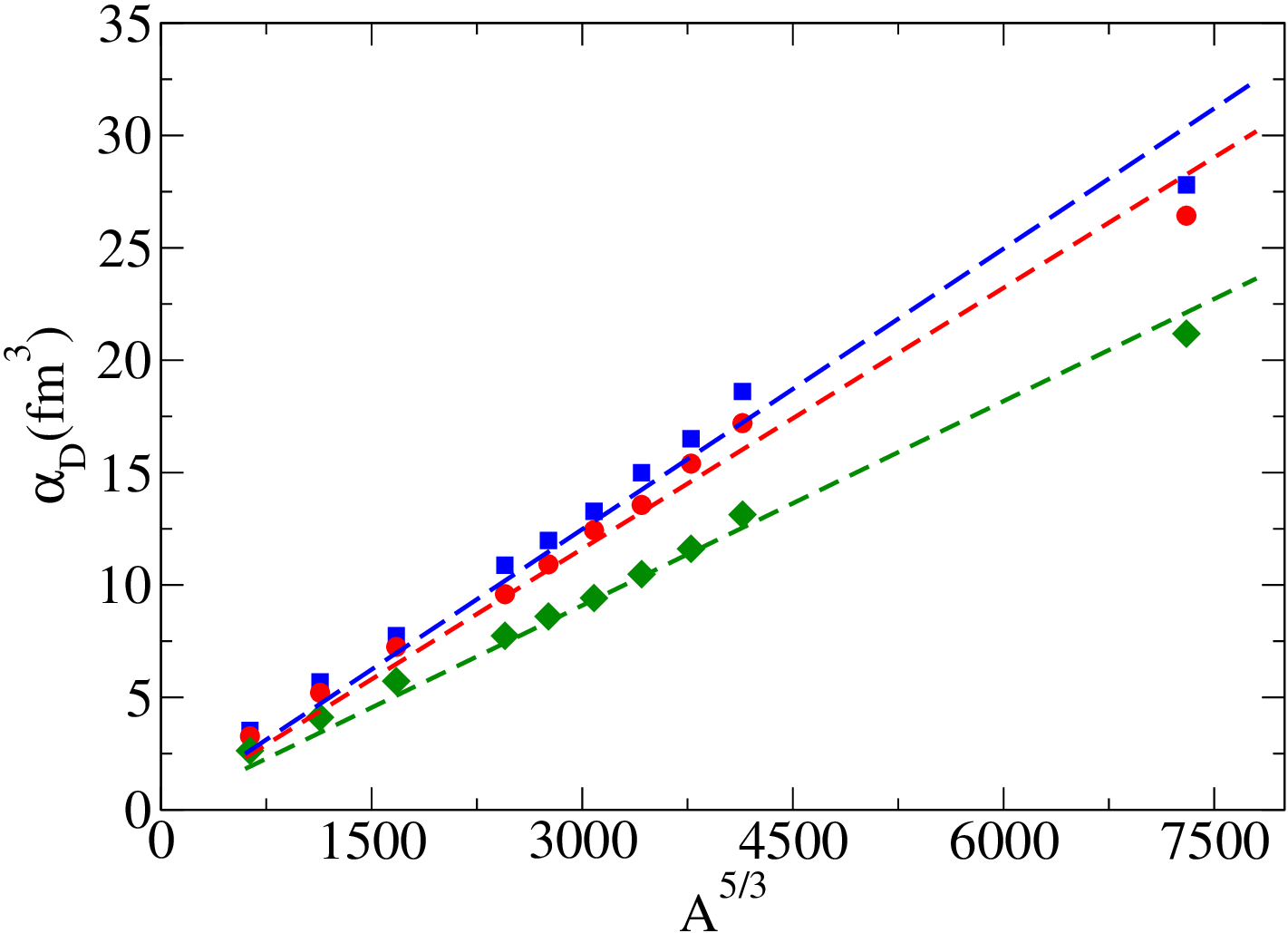}
\end{center}
\caption{(Color online) The dipole polarizability as a function of $A^{5/3}$
for asysuperstiff (blue squares) asystiff (red circles) and asysoft (green diamonds) EOS.  
The corresponding dashed lines provide the best linear fit of $\alpha_D$ with $A^{5/3}$.
The correlation coefficients $r_{fit}$ are $97\%$, $98\%$ and $99\%$ respectively.}
\label{polar}       
\end{figure}

Before discussing the dipole response below GDR region let us observe that from the strength function one can determine the 
nuclear dipole polarizability:
\begin{equation}
\alpha_D = 2 e^2 \int_0^{\infty}\frac{S(E)}{E} dE  ~.
\end{equation}
For $^{68}Ni$ the experimental value of $\displaystyle \alpha_D$ reported recently, \cite{rosPRL2013} is $3.14 fm^3$ while we
obtained values from $4.1 fm^3$ from $5.7 fm^3$ when we pass from asysoft to asysuperstiff EOS \cite{barPRC2013}. In the case of $^{208}Pb$ the 
experimental value of $\displaystyle \alpha_D$ is around $20.1 fm^3$ \cite{tamPRL2011}. In our approach it changes from 
$21.1 fm^3$ for asysoft EOS  to $28.6 fm^3$ for asysuperstiff.
Here we want to explore the mass dependence of this quantity for the three asy-EOS. In order to accomplish this goal we
consider the systems $^{48}Ca, $$^{68}Ni$, $^{86}Kr$, $^{208}Pb$ as well as the mentioned isotopic chain of Sn.
The Migdal estimation of polarizability \cite{migJP1944}, valid for large systems, provides a $A^{5/3}$ dependence with mass: 
\begin{equation}
\alpha_D = \frac{e^2 A <r^2>}{24 \epsilon_{sym}}=\frac{1.44 e^2}{40 \epsilon_{sym}} A^{\frac{5}{3}} ~,
\end{equation}
considering that $ \displaystyle  <r^2>= \frac{3}{5} R^2$ and  $\displaystyle R=1.2A^{\frac{1}{3}}$.
Since $\epsilon_{sym}$, at saturation, has similar values for the  three asy-EOS one also expect, in this situation, close values for the polarizability. 
In Fig. \ref{polar} we show the polarizability $\displaystyle \alpha_D$ as a function of $A^{5/3}$. The linear correlation is quite well
verified. Nevertheless a clear dependence of the slope with asy-EOS is evidenced. This can be related to the surface effects and the interplay
between surface and volume symmetry energy, expected to manifest in finite systems \cite{lipPLB1982} and which will be influenced by the
symmetry energy slope parameter L. 

\begin{figure}
\begin{center}
\includegraphics*[scale=0.36]{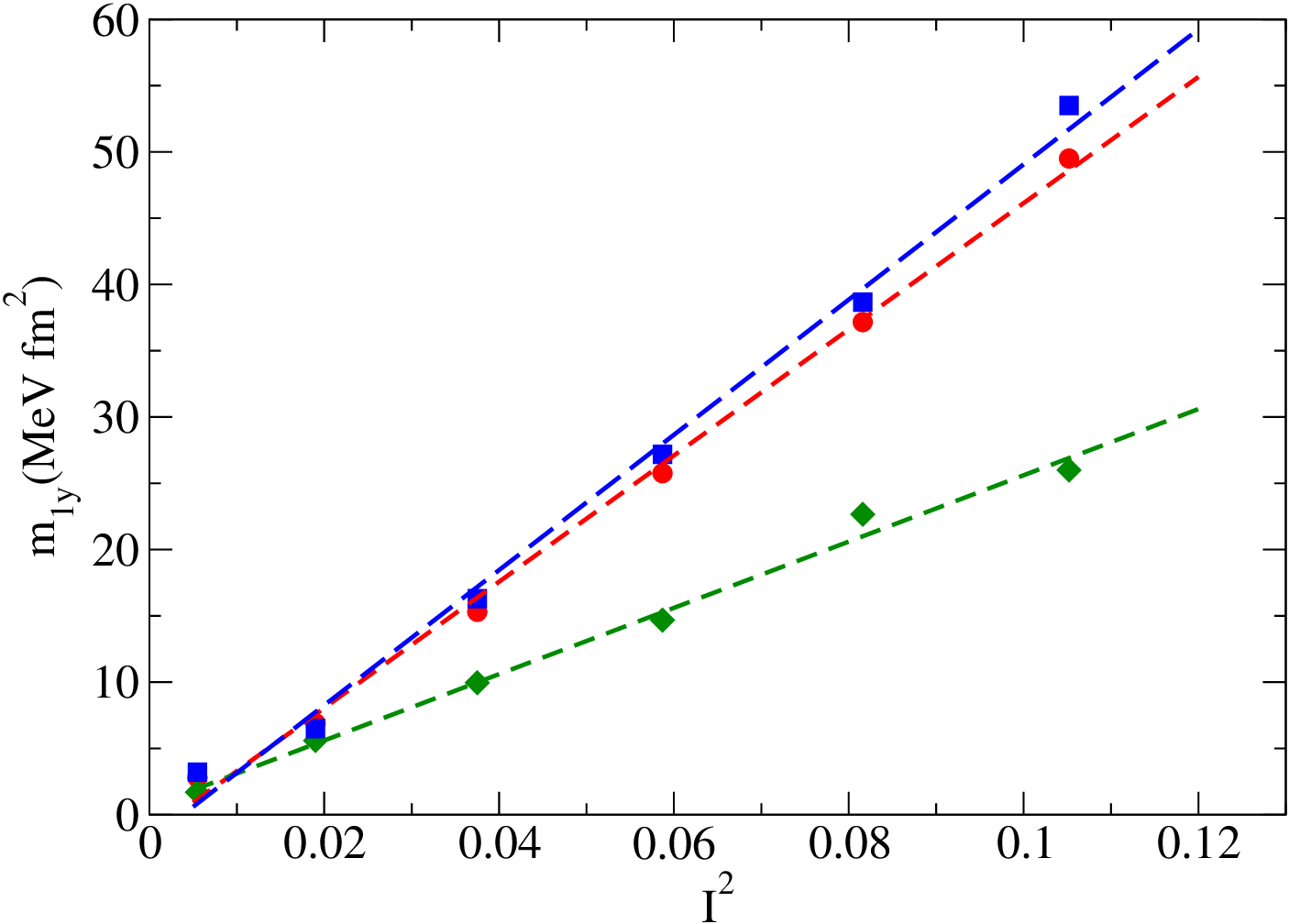}
\end{center}
\caption{(Color online) The EWSR exhausted by PDR as a function of $I$ square, for asysuperstiff (blue squares),
asystiff (red circles) and asysoft (green diamonds) EOS. Were
 considered the systems $^{108}Sn$, $^{116}Sn$, $^{124}Sn$, $^{132}Sn$, $^{140}Sn$ and $^{148}Sn$.
The dashed lines correspond to the best fit of $ m_{1,y}$ with $I^2$. The correlation coefficients
$r_{fit}$ are $99.3\%$, $99.6\%$ and $99.1\%$ respectively.}
\label{m1y}       
\end{figure}

Returning to the strength function, one can identify the appearance of a resonant response below GDR, more important when the number
of neutrons in excess is larger. In the  present model the  energy centroid  
is very well described by the parametrization $\displaystyle 41 A^{- \frac{1}{3}}$ \cite{barPRC2013}, in nice agreement with several experimental data.
This new mode we associate with Pygmy Dipole Resonance (PDR)
and notice that other studies based on Vlasov equations
arrived at similar conclusions \cite{barPRC2012,barRJP2012,abrJU2009,urbPRC2012}. 
Here our purpose is to investigate the dependence of PDR response on the isospin parameter $I$.
We calculate EWSR exhausted by this mode by integrating over the low-energy resonance region:
\begin{equation}
 m_{1,y} = \int_{PDR} E S(E) dE  ~.
\end{equation}
and plot its dependence on $\displaystyle I=\frac{N-Z}{A}$ in Fig. \ref{m1y}. 
From our calculations,  for Sn isotopes, a quadratic correlation appears to describe quite well the observed dependence of $\displaystyle m_{1y}$
with the isospin parameter $I$. We remark that, as in the case of polarization, the linear correlation between
$ m_{1,y}$ and $I^2$ is influenced by the symmetry energy slope parameter $L$. We can therefore conclude that
polarization effects in the isovector density play an important role in the dynamics of Pygmy Dipole Resonance. 

The PDR was observed experimentally for several systems \cite{aumPS2013,savPPNP2013} and discussed in different theoretical models \cite{paaJPG2010} for various nuclei, including
the Sn isotopic chain \cite{paaPLB2005,tsoPRC2008,artPRC2009,daoPRC2012,papPRC2014}.  Concerning the features of this mode, in literature still exists
an intense debate about the collective character of this mode, about the role of symmetry energy as well as about its isovector/isoscalar structure.
While the relativistic quasiparticle RPA (RQRPA) \cite{paaRPP2007,paaPRL2009} provides evidences about collectivity of PDR, from amplitudes and transition matrix elements, the nonrelativistic Hartree-Fock-Bogoliubov treatment within quasiparticle-phonon model \cite{tsoPLB2004}, assign to the resonant
structures noncollective properties. The calculations based on relativistic time-blocking \cite{litPRC2009} also shows in the dipole spectra
of even-even $\displaystyle ^{130}Sn$-$^{140}Sn$ nuclei two well separated collective structures, the lower lying one, having a specific
behavior of the transition densities of states, being ascribed to PDR. For  $\displaystyle ^{34}Mg$, from the time-dependent density plots obtained within TDHF calculations
with Skyrme interaction, was identified a superimposed surface mode, not fully coupled to the bulk dynamics. This was related to the pygmy-like peak, obtained around 10 MeV, in the dipole response strength \cite{briIJMPE2006}.

\section{Conclusions}
\label{concl}

Summarizing, the main task in this paper was to present new results regarding the collective dipole response
in connection with the properties of the symmetry energy below saturation.
Our investigation was performed in a microscopic transport model based on a system of two coupled 
Vlasov equations for protons and neutrons. 

 For all studied asy-EOS our model predicts that the energy weighted sum-rule exhausted by the
Pygmy Dipole Resonance manifests a linear dependence with the square of isospin parameter $I=(N-Z)/A$,  with a slope which
is influenced by the variation rate with density of the symmetry energy around saturation. 
Even if were considered asy-EOS providing similar values of symmetry energy at saturation
was also observed that the slope of dipole polarizability as a function of $\displaystyle A^{5/3}$
changes with the symmetry energy slope parameter L. 
We interpret these results as an indication of the surface effects associated to the polarization of 
isovector density in finite nuclei. 

In conclusion, the models based on Vlasov equation prove to be appropriate tools for the study of several
aspects of nuclear dynamics, including the development of quite feeble modes as is Pygmy Dipole Resonance, 
for which provides qualitative insights but also quantitative information regarding its dependence on the
symmetry energy or its evolution with the isospin parameter and mass number. 

\section{Acknowledgments}
This work for V. Baran and A. Croitoru was supported by a grant of the Romanian National
Authority for Scientific Research, CNCS - UEFISCDI, project number PN-II-ID-PCE-2011-3-0972.


\begin{thebibliography}{}
\bibitem{jeans1915}
J.H. Jeans, Monthly Notices Roy. Astron. Soc. \textbf{76}, (1913) 71.
\bibitem{vlasov1938}
A.A. Vlasov, Zh. Eksp. i Teor. Fiz. \textbf{8}, (1938) 291.
\bibitem{bir1991} C.K. Birdsall and A.B. Langdon, {\it Plasma Physics via Computer Simulations}, Taylor and Francis, 1991.
\bibitem{briNPA1981} D.M. Brink, M. Di Toro, Nucl. Phys. \textbf{A 372}, (1981) 151. 
\bibitem{berPR1988} G.F. Bertsch and S. Das Gupta, Phys. Rep. \textbf{160}, (1988) 189.
\bibitem{bonPR1994} A. Bonasera, F. Gulminelli, J. Molitoris, Phys. Rep. \textbf{243}, (1994) 1.
\bibitem{barPR2005} V.Baran, M. Colonna, M. Di Toro, V. Greco,
Phys. Rep. \textbf{410}, (2005) 335.
\bibitem{calAP1997} F. Calvayrac, P.G. Reinhard, E. Suraud, Ann. Phys. \textbf{225}, (1997) 125.
\bibitem{fenRMP2010} Th. Fennel et al., Rev. Mod. Phys. \textbf{82}, (2010) 1793.
\bibitem{manPRB2004} G. Manfredi, P.A. Hervieux, Phys. Rev. \textbf{B 70}, (2004) 201402.
\bibitem{croPRB2008} N. Crouseilles,  P.A. Hervieux, G. Manfredi, Phys. Rev. \textbf{B 78}, (2008) 155412.
\bibitem{moyPCPS1949} J.E. Moyal, Proc. Cambridge Phil. Soc. \textbf{45}, (1949) 99. 
\bibitem{manFIC2005} G. Manfredi, Fields Inst. Commun. \textbf{46}, (2005) 263.
\bibitem{barPRL2001} V.Baran, M. Colonna, M. Di Toro, V. Greco,
Phys. Rev. Lett. \textbf{86}, (2001) 4492.
\bibitem{larNPA1999} A.B. Larionov, M. Cabibbo, V.Baran, M. Di Toro,
Nucl. Phys. \textbf{A 648}, (1999) 157.
\bibitem{barNPA1999} V.Baran, M. Colonna, M. Di Toro, A.B. Larionov,
Nucl. Phys. \textbf{A 649}, (1999) 185C; 6th International Topical Conference
on Giant Resonances, Varenna, Italy, (1998).
\bibitem{greNPA1987} C. Gregoire et al., Nucl. Phys. \textbf{A 465}, (1987) 77.
\bibitem{schPPNP1989} P. Schuck et al., Prog. Part. Nucl. Phys. \textbf{22}, (1989) L81.
\bibitem{birPF1970} A.B. Langdon, C.K. Birdsall, The Physics of Fluids \textbf{13}, (1970) 2115.
\bibitem{idiNPA1993} D. Idier, B. Benhassine, M. Farine, B. Remaud, F. Sebille, Nucl. Phys. {\textbf A 564}, (1993) 204.
\bibitem{barPRC2013} V.Baran, M. Colonna, M. Di Toro, A. Croitoru, D. Dumitru,
Phys. Rev. \textbf{C 88}, (2013) 044610.
\bibitem{barPRC2012} V.Baran, B. Frecus, M. Colonna, M. Di Toro,
Phys. Rev. \textbf{C 85}, (2012) 051601.
\bibitem{barRJP2012} V. Baran et al., Rom. J. Phys. {\bf 57}, 36 (2012).
\bibitem{reiPRE2006} P.-G. Reinhard, P.D. Stevenson, D. Almehed, J.A. Maruhn, M.R. Strayer, Phys. Rev. {\bf E 73}, 036709 (2006).
\bibitem{abrJU2009} V.I. Abrosimov, O.I. Davydovs'ka, Ukr. J. Phys.{\bf 54}, 1068 (2009).
\bibitem{rosPRL2013} D. M. Rossi et al., Phys. Rev. Lett. \textbf{111}, (2013) 242503.
\bibitem{tamPRL2011} A. Tamii et al., Phys. Rev. Lett. \textbf{107}, (2011) 062502.
\bibitem{migJP1944} A. Migdal, J. Phys.  \textbf{8}, (1944) 331.
\bibitem{lipPLB1982} E. Lipparini, S. Stringari, Phys. Lett. {\bf B 112} (1982) 421.
\bibitem{urbPRC2012} M. Urban, Phys. Rev.{\bf C 85}, 034322 (2012).
\bibitem{aumPS2013}  T. Aumann and T. Nakamura, Phys. Scr. {\bf T 152}, 014012 (2013).
\bibitem{savPPNP2013}  D. Savran, T. Aumann and A. Zilges, Prog. Part. Nucl. Phys. {\bf 70}, (2013) 210 .
\bibitem{paaJPG2010} N. Paar, J. Phys. G: Nucl. Part. Phys. {\bf 37}, (2010) 064014.
\bibitem{paaPLB2005} N. Paar, T. Niksic ,D. Vretenar, P. Ring, Phys. Lett. {\bf B 606} (2005) 288.
\bibitem{tsoPRC2008} N. Tsoneva, H. Lenske,  Phys. Rev. {\bf C 77} (2008) 024321.
\bibitem{artPRC2009} D.P. Arteaga, E. Khan, P. Ring,  Phys. Rev. {\bf C 79} (2009) 034311.
\bibitem{daoPRC2012} I. Daoutidis, S. Goriely,  Phys. Rev. {\bf C 86} (2012) 034328.
\bibitem{papPRC2014} P. Papakonstantinou, H. Hergert, V. Yu. Ponomarev, R.Roth,  Phys. Rev. {\bf C 89} (2014) 034306.
\bibitem{paaRPP2007} N. Paar, D. Vretenar, E. Khan, G. Colo, Rep. Prog. Phys. {\bf 70}, (2007) 691.
\bibitem{paaPRL2009} N. Paar, Y.F. Niu, D. Vretenar, J. Meng, Phys. Rev. Lett. \textbf{103}, (2009) 032502.
\bibitem{tsoPLB2004} N. Tsoneva, H. Lenske, Ch. Stoyanov,  Phys. Lett. {\bf B 586} (2004) 213.
\bibitem{litPRC2009} E. Litvinova, P. Ring, V. Tselyaev, K. Langanke, Phys. Rev.{\bf C 79}, (2009) 054312.
\bibitem{briIJMPE2006} M.P. Brine, P.D. Stevenson, J.A. Mahrun, P.-G. Reinhard,  Int. J. Mod. Phys. {\bf E 15} (2006) 1417.
\end{thebibliography}
\end{document}